\documentclass[12pt]{article}
\usepackage[T1]{fontenc}
\usepackage[utf8]{inputenc}
\usepackage{lmodern}
\usepackage{amsmath}
\usepackage{amssymb}
\usepackage{graphicx}
\usepackage{caption}
\usepackage{subcaption}
\usepackage{multirow}
\usepackage{float}

\title{The far field limit profile functions of two overlapped dyons}
\author{Motahareh Kiamari$^{1}$, Sedigheh Deldar $^{1}$ \\
$^1$Department of Physics, University of Tehran,\\
\small P.O. Box 14395/547, Tehran 1439955961,
Iran.}
\date{}
\begin{document}

\maketitle

\begin{abstract}
We study the profile functions of two overlapped dyons at 
infinity. We assume that the superposition of two dyons satisfies 
Yang-Mills (Y-M) equation and then we find new equations of motion for 
individual dyons which do not satisfy the original Y-M equation, 
anymore. By solving these new equations, we find out that 
the profile functions of two overlapped dyons of the same type at infinity, looks like the profile functions of one 
dyon. However, the superposition of two dyons of different 
types gives trivial holonomy and therefore no contribution in the 
confinement phase is observed.
\end{abstract}

\section{Introduction}
\label{introduction}

Calorons as the candidates of the vacuum structure of QCD, have been of 
interest since the time they were found independently by Kraan and van Baal \cite{Kraan98}, 
and Lee and Lu \cite{Lee98} in 1998. These structures are the 
periodic solutions of Yang-Mills equations and have non-Abelian and 
non-linear structures, hence their interactions are complex. There have 
been some attempts to study the interaction of calorons or to construct 
the multi-caloron configurations \cite{Bruck2002}, \cite{Ger2007}.

A common method to discuss about calorons is to focus on dyons, the 
constituents of calorons. Dyons are the static and self-dual solutions 
of Y-M equations with electric and magnetic color charges. Although 
dyons have non-linear and non-Abelian structures in their cores, they 
have a simpler structure than calorons. Since dyons, just like calorons, 
are non-Abelian structures, due to the 
non-Abelian part of the strength tensor of $SU(N)$ gauge group, the superposition of two or more dyons is not the appropriate solution for the Y-M equations. 
The usual method to deal with this problem is to transform dyons to the gauge in 
which their superposition is as close as possible to the solution of Y-M 
equations. Since dyons are formulized in the hedgehog gauge, one can 
transform them to the stringy gauge in which dyons are Abelian at 
infinity and the superposition of two or more dyons is the solution of 
Y-M equations, at least at infinity \cite{Mart95}. It is worth emphasizing 
that, we can add dyons in all gauges, and the only reason which 
restricts us to the stringy gauge is that we want the superposition of 
dyons to be the solution of Y-M equations.

There is another well known method to deal with non-linear objects which is mostly used by people in Optics.
 It is assumed that the superposition of the objects is the solution of the equations of motion and then modify the equations for each object. 
Therefore, the individual object satisfies the modified equation, not 
the original one \cite{Agr2007}. In our previous paper \cite{KiaDel2018}, we have already applied this 
procedure to overlapped dyons and have calculated the near field limit profile functions. We have assumed the superposition of two dyons in the 
hedgehog gauge or Rossi gauge \cite{Ros82} as the solution of Y-M equations 
and then have found the "perturbed" version of equation for dyons. We called it perturbed, since we did not change the direction of the gauge 
fields of the individual dyons and just modified their profile functions 
with some approximations. We found the profile functions of the 
perturbed dyons near their centers when their cores were overlapped 
\cite{KiaDel2018}. 

In this paper, we find the perturbed equations for the far field limit to 
solve them for large distances with respect to the centers of the dyons and we study the 
effects of the presence of one dyon on the other one, at large distances. 
Then, we study the electric and magnetic fields, charges, the 
topological charge and the action for each dyon at the presence of another one for two kinds of dyons in $SU(2)$ gauge group. We find that the superposition of two 
dyons of the same type leads to the observation of just one dyon, at 
infinity. However, the superposition of two dyons of different types has 
trivial holonomy and has no contribution in the confinement phase.

 The paper is organized as the following. In section \ref{sec:dyon}, the important 
properties of dyons are briefly introduced. In section \ref{sec:calorons}, a short review of the properties of calorons and their superposition is represented. In section \ref{sec:superposition}, we find the 
equations of motion for perturbed dyons and solve them for the given boundary 
conditions. In section \ref{sec:MM,LL}, we focus on the superposition of two dyons of the same type and 
calculate some of its important properties. In section \ref{sec:ML}, we study the 
superposition of two dyons of different types as the solution of Y-M 
equation. We conclude the paper in section \ref{sec:concl}.

\section{Dyons in $SU(2)$ Yang-Mills theory}
\label{sec:dyon}
Dyons are basically $SU(2)$ configurations which can be 
generalized to $SU(N)$ gauge group. In $SU(2)$, there are two kinds of 
(anti-) self-dual dyon based on (anti-) self-dual equations of motion,
\begin{equation}
  F_{\mu \nu }^{a}=\pm {\widetilde{F}}_{\mu \nu }^{a} \rightarrow  E_{i}^{a}=\pm B_{i}^{a},
  \label{self-dual-eq}
\end{equation}
where the plus and minus signs represent self-dual and anti-self-dual 
equations, respectively. $F$ is the strength tensor of $SU(2)$ gauge group 
and $E$ and $B$ are the electric and magnetic fields,
\begin{equation}
  F_{\mu \nu }^{a}=\partial _{\mu }A_{\nu }^{a}-\partial _{\nu }A_{\mu }^{a}+\epsilon _{abc}A_{\mu }^{b}A_{\nu }^{c},
\end{equation}
\begin{equation}
  E_{i}^{a}=F_{i4}^{a},
  \label{Ei}
\end{equation}
\begin{equation}
  B_{i}^{a}=\frac{1}{2}\epsilon _{ijk}F_{jk}^{a}.
  \label{Bi}
\end{equation}
In this paper, we focus on self-dual equation and dyons, but all we do 
can be repeated for anti-self-dual equation, straightforwardly. The 
first dyon called M, is defined by the gauge fields in the hedgehog 
gauge,
\begin{equation}
  A_{4}^{a}(r)=n_{a}\frac{E(r)}{r}, \;\;\;\; E(r)=1-\nu r\coth \nu r,
  \label{A4}
\end{equation}
\begin{equation}
  A_{i}^{a}(r)=\epsilon _{aij}n_{j}\frac{1-A(r)}{r}, \;\;\;\; A(r)=\frac{\nu r}{\sinh \nu r},
  \label{Ai}
\end{equation}
with electric and magnetic charges (+,+) and the second dyon is called L 
with charges (-,-) and is obtained by the replacement of $\nu \rightarrow  
2\pi T-\nu $, where $\nu \equiv \sqrt{A_{4}^{a}A_{4}^{a}}\vert _{r\rightarrow  
\infty }$ is called the holonomy. It is the order parameter which 
specifies the confinement-deconfinement transition phase. 

The profile functions $E(r)$ and $A(r)$ are properly satisfied the 
boundary conditions, as will be explained. To avoid singularity at the 
origin,
\begin{equation}
  \lim _{r\rightarrow  0}E(r)=0, \;\;\;\; \lim _{r\rightarrow  0}A(r)=1,
  \label{bc0}
\end{equation}
and to have a finite action,
\begin{equation}
  \lim _{r\rightarrow  \infty }E(r)=-\nu r+1, \;\;\;\; \lim _{r\rightarrow  \infty }A(r)=0.
  \label{bc-inf}
\end{equation}
With these boundary conditions, the electric and magnetic fields become 
coulomb-like and the electric and magnetic charges are well-defined. 
Hence, one can gauge comb a dyon to the "stringy gauge" at infinity,
\begin{equation}
  A_{i}^{a}\xrightarrow{r\rightarrow  \infty }0, \;\;\;\;
A_{4}^{a}\xrightarrow{r\rightarrow  \infty }-\nu \frac{\sigma ^{3}}{2},
\end{equation}
where $\sigma ^{3}$ is the third Pauli matrix. The topological charge 
of YM fields is defined by
\begin{equation}
  Q_{T}=\frac{1}{32\pi ^{2}}\int{d^{4}}x F_{\mu \nu }^{a}{\widetilde{F}}_{\mu \nu}^{a}=-\frac{1}{8\pi ^{2}}\int{d^{4}}xE_{i}^{a}B_{i}^{a},
\end{equation}
By substitution of gauge fields of dyon in Eqs. (\ref{Ei}) and (\ref{Bi}), $E_{i}^{a}$ and $B_{i}^{a}$ is obtained,
\begin{equation}
  E_{i}^{a}=(\delta_{ai}-n_{a}n_{i})\frac{A(r)}{r}\frac{E(r)}{r}+n_{a}n_{i}\frac{d}{dr}\left( \frac{E(r)}{r}\right),
  \label{new-Ei}
\end{equation}
\begin{equation}
  B_{i}^{a}=(\delta_{ai}-n_{a}n_{i})\frac{1}{r}\frac{dA(r)}{dr}+n_{a}n_{i}\frac{A^{2}(r)-1}{r^{2}}, 
  \label{new-Bi}
\end{equation}
Then, $Q_{T}$ is obtained,
\begin{equation}
\begin{split}
  Q_{T}&=-\frac{1}{8\pi ^{2}}\int_{0}^{\frac{1}{T}}{dx_{4}\int{d^{3}}rE_{i}^{a}B_{i}^{a}}\\{}
  &=-\frac{1}{2\pi T}\int{dr\frac{d}{dr}\frac{(1-A^{2})E}{r}}=-\frac{1}{2\pi T}\frac{(1-A^{2})E}{r}\vert _{0}^{\infty }=\frac{\nu }{2\pi T}, \\{}
  \end{split}
  \label{tcharge}
\end{equation}
where the topological charge is $1/2$ for confinement phase where $\nu =\pi 
T$, for both M and L dyons. The action of the self-dual fields is
\begin{equation}
  S=\frac{1}{4g^{2}}\int{d^{4}}x F_{\mu \nu }^{a}F_{\mu \nu }^{a}=
  \frac{1}{4g^{2}}\int{d^{4}}x F_{\mu \nu }^{a}{\widetilde{F}}_{\mu \nu }^{a}=\frac{8\pi ^{2}}{g^{2}}Q_{T}, 
  \label{action}
\end{equation}
which is the same for both kind of dyons in confinement phase.

\section{Review of calorons and their superposition}
\label{sec:calorons}
In 1998, Kraan and van Baal \cite{Kraan98}, as well as Lee and Lu \cite{Lee98} independently introduced the periodic instanton or caloron, as an exact solution of the Y-M equations with a non-trivial holonomy. Kraan and van Baal constructed the caloron as an infinite, periodic chain of instantons, using the Nahm transformation and ADHM techniques. 
\\This new structure has some interesting features. For small $\rho$, where $\rho$ is the scale parameter or radius of this structure, a caloron approaches to a simple instanton. As $\rho$ increases, two lumps are identified in a caloron. For large $\rho$, these two lumps are well-separated and as $\rho \rightarrow \infty$ they will be static and they will have spherical symmetry far away from their centers. Since they are static and self-dual, these lumps are BPS monopoles or dyons. Therefore, for large $\rho$, a dyonic picture is more suitable for describing calorons. In this picture, a caloron in $SU(2)$ gauge group consists of two M and L dyons where the dyon L is twisted with a time-dependent gauge transformation in the direction of holonomy. The stability of the caloron \cite{Kraan98} is guarantied by this "twisting procedure". 
\\Near the dyon cores, the temporal component of the gauge fields, $A_{4}$, are in hedgehog gauge, while far from the dyons, a caloron is in unitary gauge with diagonal $A_{4}$ and a singular Dirac string is located along the line connecting the two dyons \cite{Kraan98}\cite{Bruck2010}. Hence, a caloron is not the superposition of two original dyons even for large $\rho$.
\\Considering the above discussions, if one wants to study the interaction of calorons, the dyonic picture is simpler, since the interaction of calorons is reduced to the interaction of dyons plus the interaction of dyon with a Dirac string.
Gerhold \textit{et. al.} studied the interaction of two calorons for large $\rho$. They showed that if an object is  located on top of the Dirac string of a caloron, the superposition of the caloron and that object is the sum of the unchanged field of the caloron and the gauge rotated field of that object \cite{Ger2007},
\[ A_{\mu}(x) = A_{\mu}^{caloron}(x)+e^{-iG(x)\tau_{3}}.A_{\mu}^{object}.e^{iG(x)\tau_{3}}. \]
They also used pseudo-ADHM superposition scheme for a combination of calorons and compared their results with sum ansatz results \cite{Ger2007}.
\\In this paper and our previous one \cite{KiaDel2018}, we have planned to find a solution for the superposition of calorons or dyons. What we are doing is similar to what  Gerhold \textit{et. al.} did \cite{Ger2007}, but not to construct any new stable configuration like calorons. For this purpose, we put calorons close to each other at large $\rho$, in such a way that the cores of their dyons are overlapped. Then, we study the interactions of the same charged dyons as well as the different charged dyons. Thinking of caloron as the only way to make interaction between two dyons of the opposite charges is limitative and these structures, dyons, can interact with each other just like other structures, for example vortices, either as the independent objects in the dyonic ensemble or as the constituents of calorons for large $\rho$. 
\\Each dyon is an Abelian and linear objects outside its core, while it is an $ SU(2) $ object and hence non-Abelian in its core. The superposition of two self-dual dyons is not the solution of Y-M equations, because of their non-linearity nature. We assume that the sum ansatz of the two dyons is the solution of Y-M equations and then we find new equations for each dyons, and we call them “deformed” dyons. This is a well-known procedure to study the interaction of the non-linear objects. It is obvious that the deformed dyon is not the solution of Y-M equations anymore and hence is not a $ U(1) $ object at infinity and does not have a well-defined charges.
\\As mentioned before, dyons are in the hedgehog gauge near their cores. Hence, we study the superposition of two dyons in this gauge as will be explained in details in section \ref{sec:superposition}. In all of our calculations, the cores of dyons are overlapped, but by increasing the distance between the centers of the dyons in such a way that the dyons become well-separated, one can gauge comb the dyons to the stringy gauge and consider them as $U(1)$ objects with well-defined electric and magnetic charges and study their interactions \cite{Manton}.

Our other motivation to study the overlapped dyons, is to improve the behavior of dyonic ensembles in our previous papers. In \cite{KiaDel2017}, we have studied a dyon gas by Particle Mesh Ewald’s (PME) method to calculate the free energy of quark-antiquark pair in the confinement phase. Based on the paper by Bruckmann  \textit{et. al.} \cite{Bruck12}, we initially considered the non-interacting ensemble of dyons. In that paper, dyons were in an appropriate separations, so that they could be considered as $U(1)$ particles with Coulombic electric and magnetic fields. Then, we added the metric of moduli space as an effective interaction \cite{KiaDel17} and recalculated the free energy. We may use the result of this new research to modify the free energy of the ensemble of dyons.


\section{Profile functions of two overlapped dyons in the far field limit}
\label{sec:superposition}
Recently, we have studied the superposition of two dyons when their 
cores overlap \cite{KiaDel2018}. We followed the well-known procedure in optics to 
study the interaction of non-linear objects \cite{Agr2007}. Dyons in $SU(2)$ 
gauge group are non-Abelian and non-linear. However, they become Abelian 
and can be gauge combed to the preferred direction at infinity where 
they obtain finite and well-defined action. Since dyons are non-Abelian, 
their superposition is not the solution of Y-M equation. To have a 
multi-dyon configuration that is the solution of Y-M equation, we do a 
simple assumption. We assume the superposition of two dyons is the 
solution of Y-M equations and find new equations of motion for each 
individual dyon. In other words, we assume that the two dyons deform each 
other in such a way that their superposition becomes the solution of the Y-M 
equation. It means that the presence of one dyon on another one has the 
effect of changing the original Y-M equations to a modified one for each of those dyons. 

\begin{figure}
 \captionsetup{font=footnotesize}
  \begin{center}
    \includegraphics[width=0.6\linewidth]{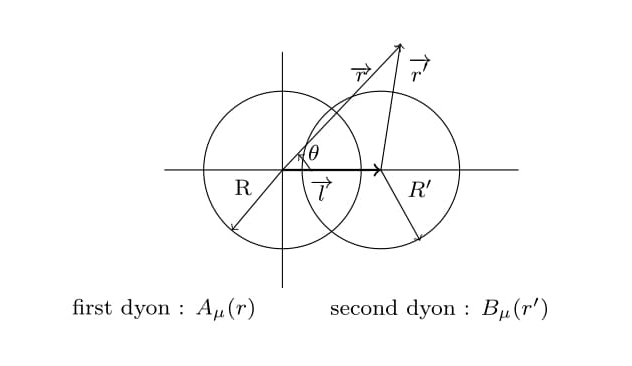}
    \caption{Two-dyon configuration. The first dyon is located at the origin and defined by the gauge fields $A_{\mu}(r)$, while the second dyon is located by a distance $l$ from the first one and defined by the gauge fields $B_{\mu}(r')$}
       \label{fig:2dyons}
  \end{center}
\end{figure}

Consider two dyons in the hedgehog or Rossi gauge where the first one is 
located at the origin and associated by the gauge fields $A_{\mu }(r)$ 
and the second one is separated by a distance $l\equiv \vert 
\overrightarrow{l}\vert $ from the first one and is associated by the 
gauge fields $B_{\mu }(r')$, Fig. (\ref{fig:2dyons}). 

As mentioned before, we can consider dyons in any gauge we need, since 
the sum ansatz is assumed to be the solution of Y-M equations in all 
spatial space. Hence, we do not have to transform dyons to the stringy 
gauge to be as close as possible to the solution of Y-M equations at 
infinity. Instead, we assume the sum ansatz is the solution of Y-M 
equation.

Assume $A_{\mu }(r)+B_{\mu }(r')$ is the solution of the 
self-dual equation (\ref{self-dual-eq}) in all spatial space,
\[   A_{\mu }^{\prime a}=A_{\mu }^{a}+B_{\mu }^{a}, \]
substituting it to Eq. (\ref{self-dual-eq}) and using the definition of the electric and 
magnetic fields of Eqs. (\ref{Ei}) and (\ref{Bi}), we have
\begin{equation}
 \partial _{i} A^{\prime a}_{4} - \partial _{4}A^{\prime a}_{i} + \epsilon _{abc} A^{\prime b}_{i} A^{\prime c}_{4} - \frac{1}{2} \epsilon _{ijk} \left( \partial _{j} A^{\prime a}_{k} - \partial _{k} A^{\prime a}_{j} + \epsilon _{abc} A^{\prime b}_{j} A^{\prime c}_{k} \right) = 0.
\end{equation}
and after some calculations,
\begin{equation}
\resizebox{\textwidth}{!}{$F^{a}_{i4} - \frac{1}{2}\epsilon _{ijk} F^{a}_{jk} + \epsilon_{abc} A^{b}_{i} B^{c}_{4} - \frac{1}{2} \epsilon _{ijk} \epsilon _{abc} A^{b}_{j}B^{c}_{k} = - \left[ G^{a}_{i4} - \frac{1}{2}\epsilon _{ijk} G^{a}_{jk} + \epsilon_{abc} B^{b}_{i} A^{c}_{4} - \frac{1}{2} \epsilon _{ijk} \epsilon _{abc} B^{b}_{j}A^{c}_{k} \right].$}
\label{fg}
\end{equation}
where $F$ and $G$ are the strength tensors of the first and the second dyons, 
respectively. Since Eq. (\ref{fg}) is satisfied for all values of functions $A$ and $B$ 
 and for all values of $r$ and $r'$, the two sides of 
equations should be equal to a constant that we choose it to be zero. 
Hence the first dyon satisfies the modified version of Y-M equation,
\begin{equation}
  F_{i4}^{a}-\frac{1}{2}\epsilon _{ijk}F_{jk}^{a}=-\epsilon _{abc}A_{i}^{b}B_{4}^{c}
  +\frac{1}{2}\epsilon _{ijk}\epsilon _{abc}A_{j}^{b}B_{k}^{c},
  \label{perturbed-eq}
\end{equation}
not the original self-dual equation,
\begin{equation}
  E_{i}^{a}=B_{i}^{a} \rightarrow  F_{i4}^{a}-\frac{1}{2}\epsilon_{ijk}F_{jk}^{a}=0.
\end{equation}
We call the extra term in Eq. (\ref{perturbed-eq}) "perturbation term" which causes 
some kind of deformation in the first dyon and an interaction between two dyons. 
Calculating two sides of Eq. (\ref{perturbed-eq}), 
\begin{equation}
  LHS=(\delta_{ai}-n_{a}n_{i})\left( \frac{A_{p}(r)}{r}\frac{E_{p}(r)}{r}-\frac{1}{r}\frac{dA_{p}(r)}{dr}\right)+n_{a}n_{i}\left(\frac{d}{dr}\left(\frac{E_{p}(r)}{r}\right)-\frac{A_{p}^{2}(r)-1}{r^{2}}\right),
\end{equation}
\begin{equation}
  RHS=(\delta_{ai}-n_{a}n_{i})\frac{1-A(r)}{r}\frac{E(r')}{r'}+n_{a}n_{i}\frac{1-A(r)}{r}\frac{1-A(r')}{r'},
\end{equation}
we obtain new equations for the deformed dyons to solve,
\begin{equation}
  A_{p}(r)\frac{E_{p}(r)}{r}-\frac{d}{dr}A_{p}(r)=(1-A(r))\frac{E(r^{'})}{r^{'}},
  \label{new1}
\end{equation}
\begin{equation}
  r\frac{d}{dr}\left(\frac{E_{p}(r)}{r}\right)-\frac{A_{p}^{2}(r)-1}{r}=(1-A(r))\frac{1-A(r^{'})}{r^{'}},
  \label{new2} 
\end{equation}
where the subscript $p$ denotes the perturbed version of the $A(r)$ and $E(r)$ of Eqs. (\ref{A4}) and (\ref{Ai}). In the previous paper \cite{KiaDel2018}, we solved Eqs. (\ref{new1}) and (\ref{new2}) inside the 
cores of dyons $(r\rightarrow  0)$, for $l\rightarrow  0$ to make sure that the cores 
of dyons overlap. We approximated the right hand side of Eqs. (\ref{new1}) and 
(\ref{new2}) by the asymptotic values of profile functions of individual dyons 
at the origin. We also approximated the direction of the gauge fields of the 
second dyon by the direction of the first one, $n'_{a}\approx n_{a}
$. Considering $l\rightarrow 0$, this approximation is good enough for 
non-overlapping region of the cores and just acceptable for overlapping 
region. We found that the presence of one dyon affects the profile 
functions of another one and they depend on the polar angle, $\theta $ 
in Fig. (\ref{fig:2dyons}), and the spherical symmetry is lost \cite{KiaDel2018}.

In this paper, we solve Eqs. (\ref{new1}) and (\ref{new2}) outside the cores of dyons, 
\textit{i.e.} at infinity. Since we do our calculations in 
infinity, our approximations $l\rightarrow  0$, $n'_{a}\approx n_{a}$ and $r'\approx r$ are reasonable,
\begin{equation}
  A(r)\frac{E(r)}{r}-\frac{d}{dr}A(r)\approx (1-A(r))\frac{E(r)}{r},
  \label{app-new1}
\end{equation}
\begin{equation}
  r\frac{d}{dr}\left(\frac{E(r)}{r}\right)-\frac{A^{2}(r)-1}{r}\approx (1-A(r))\frac{1-A(r)}{r}.
  \label{app-new2}
\end{equation}
Instead of asymptotic values for profile functions of individual dyons at 
infinity, we put the exact functions on the right hand sides of Eqs. 
(\ref{app-new1}) and (\ref{app-new2}) and then, the subscript $p$ can be omitted. We suggest the solution of a deformed dyon just like the 
profile functions of an individual dyon of Eq. (\ref{bc-inf}), 
\begin{equation}
  E(r)=ar+b,  \;\;\;\;  A(r)=c,
  \label{solution}
\end{equation}
This is done because we prefer to save the structure of the original dyon and make the slightest 
changes on the dyon in order to make the superposition of two dyons as 
the solution of Y-M equation. For two dyons of the same kind $
c=b=\frac{1}{2}$ and $ a$ is the free parameter. This solution should 
satisfy the boundary condition at infinity to have finite action. 
However, $F_{\mu \nu }$ must tend to zero faster than or equal to $
\frac{1}{r^{2}}$ as $r\rightarrow  \infty $. Hence, we should calculate the 
electric (magnetic) field of the superposition of two dyons to make sure 
that our solution satisfies the boundary conditions.
\begin{equation}
  A_{\mu }^{\prime a}=A_{\mu }^{a}+B_{\mu }^{a} \to F_{\mu \nu }^{\prime a}=F_{\mu \nu }^{a}+G_{\mu \nu }^{a}+H_{\mu \nu }^{a},
\end{equation}
and 
\begin{equation}
  H_{\mu \nu }^{a}=\epsilon _{abc}(A_{\mu }^{b}B_{\nu }^{c}+B_{\mu }^{b}A_{\nu }^{c}),
\end{equation}
where $H_{\mu \nu }^{a}$ is the extra term due to the non-Abelianity 
of $SU(2)$ gauge group which leads to the interaction term in action. The 
electric field of the superposition $A'$ is,
\begin{equation}
  E_{i}^{\prime a}= E_{i}^{(1)a}+E_{i}^{(2)a}+H_{i4}^{a},
\end{equation}
where the indices (1) and (2) point to the first and second dyons, 
respectively and $E_{i}^{(1)a}=F_{i4}^{a}$ and $ 
E_{i}^{(2)a}=G_{i4}^{a}$, which is calculated in Eq. (\ref{new-Ei}). Since we 
consider the approximations $n'_{a}\approx n_{a}$ and $r'\approx r$ at infinity, the electric fields of both dyons are equal,
\begin{equation}
  E_{i}^{(1)a}+E_{i}^{(2)a}=2(\delta_{ai}-n_{a}n_{i})\frac{A(r)}{r}\frac{E(r)}{r}+2n_{a}n_{i}\frac{d}{dr}\left(\frac{E(r)}{r}\right),
\end{equation}
\begin{equation}
\begin{split}
  H_{i4}^{a}&=\epsilon_{abc}(A_{i}^{b}B_{4}^{c}+B_{i}^{b}A_{4}^{c}) \\{}
  &=\epsilon _{abc}\left[
\left(\epsilon_{bij}n_{j}\frac{1-A(r)}{r}\right)\left(n'_{c}\frac{E(r')}{r'}\right)+\left(\epsilon_{bij}n'_{j}\frac{1-A(r')}{r'}\right)\left(n_{c}\frac{E(r)}{r}\right)\right] \\{}
&\approx 2\epsilon _{abc}\left(\epsilon_{bij}n_{j}\frac{1-A(r)}{r}\right)\left(n_{c}\frac{E(r)}{r}\right)=-2(\delta_{ai}-n_{a}n_{i})\frac{1-A(r)}{r}\frac{E(r)}{r},
\end{split}
\end{equation}
then
\begin{equation}
  E_{i}^{\prime a}=2(\delta_{ai}-n_{a}n_{i})\left(\frac{A(r)}{r}\frac{E(r)}{r}-\frac{1-A(r)}{r}\frac{E(r)}{r}\right)+2n_{a}n_{i}\frac{d}{dr}\left(\frac{E(r)}{r}\right),
  \label{new-E'}
\end{equation}
substituting the ansatz of Eq. (\ref{solution}) into Eq. (\ref{new-E'})
\begin{equation}
\begin{split}
  E_{i}^{\prime a}& =2(\delta_{ai}-n_{a}n_{i})\left(\frac{c}{r}\frac{ar+b}{r}-\frac{1-c}{r}\frac{ar+b}{r}\right)+2n_{a}n_{i}\frac{d}{dr}\left(\frac{ar+b}{r}\right) \\{}
&\xrightarrow{c=b=\frac{1}{2}} E_{i}^{\prime a}=-\frac{n_{a}n_{i}}{r^{2}}. 
\end{split}
\end{equation}
Since $A'$ is the self-dual solution, $E_{i}^{a}=B_{i}^{a}$ and the 
magnetic field is the same. However, the calculation of the magnetic 
field is straightforward. Therefore, the electric and magnetic fields of 
the superposition of two dyons are $\frac{1}{r^{2}}$ at infinity and 
our solution satisfies the boundary conditions.

\section{The superposition of Two Dyons of the Same type}
\label{sec:MM,LL}

Now, we study the gauge fields of the superposition of two dyons $
A'$ with more details. 
\begin{equation}
  A_{\mu }^{\prime a}(r)=A_{\mu }^{a}(r)+B_{\mu }^{a}(r'),
  \label{A'}
\end{equation}
where from Eqs. (\ref{A4}) and (\ref{Ai}),
\begin{equation}
  A_{4}^{\prime a}(r)=A_{4}^{a}(r)+B_{4}^{a}(r')=n_{a}\frac{E(r)}{r}+n'_{a}\frac{E(r')}{r'}, 
  \label{new-A4}
\end{equation}
\begin{equation}
  A_{i}^{\prime a}(r)=A_{i}^{a}(r)+B_{i}^{a}(r')=\epsilon_{aij}n_{j}\frac{1-A(r)}{r}+\epsilon _{aij}n'_{j}\frac{1-A(r')}{r'}. 
  \label{new-Ai}
\end{equation}
At infinity, $n'_{a}\approx n_{a}$ and $r'\approx r$ are 
reasonable approximations and by the ansatz of Eq. (\ref{solution}), the 
superposition of two dyons of the same type is defined,
\begin{equation}
\begin{split}
  & A_{4}^{\prime a}(r)\approx 2n_{a}\frac{E(r)}{r}=n_{a}\left(2a+\frac{1}{r} \right), \\{}
& A_{i}^{\prime a}(r)\approx 2\epsilon _{aij}n_{j}\frac{1-A(r)}{r}=\epsilon _{aij}n_{j}\frac{1}{r}.
\end{split}
\end{equation}
Now, we can consider the superposition of two dyons of the same type as 
one dyonic structure and find the profile functions of this solution of 
Y-M equation,
\begin{equation}
  A_{4}^{\prime a}(r)=n_{a}\frac{E'(r)}{r}=n_{a}\left(2a+\frac{1}{r} \right) \rightarrow  E'(r)=2ar+1,
\end{equation}
\begin{equation}
  A_{i}^{\prime a}(r)=\epsilon _{aij}n_{j}\frac{1-A'(r)}{r}=\epsilon _{aij}n_{j}\frac{1}{r} \rightarrow  A'(r)=0.
\end{equation}
Comparing with the asymptotic values of the profile functions of an 
individual dyon of Eq. (\ref{bc-inf}), we can choose $a=-\nu /2$. We recall that $a$ is 
a free parameter. Then, the profile functions of the superposition of 
two dyons, $A'$, are the same as the profile functions of an original 
individual dyon. In other words, the superposition of two M(L) dyons 
behaves like one M(L) dyon at infinity, if it is forced to be the 
solution of Y-M equations. As mentioned before, L dyon can be obtained 
from the gauge fields of the M dyon just by the replacement $\nu \rightarrow  
2\pi T-\nu $.

To define the Abelian and well-defined electric and magnetic charges and 
the holonomy at infinity, one should rotate the electric and magnetic 
fields and $A_{4}^{'a}$ along the third color axis, $\sigma ^{3}$. 
This can be done by the help of an unitary matrix $S_{-}(\theta 
,\varphi )$,
\begin{equation}
  S_{-}(\theta, \varphi )= e^{i\frac{\varphi }{2}\sigma 
^{3}}e^{i\frac{\pi -\theta }{2}\sigma ^{2}}e^{i\frac{\varphi }{2}\sigma ^{3}} 
\rightarrow  S_{-}(n.\sigma )S_{-}^{\dagger}=-\sigma ^{3},
\end{equation}
Then the $A_{4}$ component of the superposition of two M dyons is, 
\begin{equation}
  A_{4}^{MM}=\left(\nu -\frac{1}{r}\right)\frac{\sigma ^{3}}{2},\;\; r\rightarrow  \infty
\end{equation}
and the radial component of the electric and magnetic fields of the 
superposition of two M dyons at infinity are coulomb-like and Abelian
\begin{equation}
  E_{r}^{MM}=B_{r}^{MM}=\frac{1}{r^{2}}\frac{\sigma ^{3}}{2},\;\; r\rightarrow  \infty
\end{equation}
while the $\theta, \varphi $ components of these fields are non-zero 
only inside the cores of dyons. Therefore, the superposition of two M 
dyons has Abelian electric and magnetic charges (+,+). In other 
words, when we look at the superposition of two M dyons from infinity, 
we see the point-like charge with electric and magnetic charges (+,+), 
just like the original M dyon. The whole calculations can be repeated 
for the superposition of two L dyons, except the electric and magnetic 
fields and $A_{4}^{\prime a}$ should be rotated by $S_{+}(\theta, \varphi )$ \cite{Dia2009},
\begin{equation}
  S_{+}(\theta, \varphi )= e^{-i\frac{\varphi }{2}\sigma ^{3}}e^{i\frac{\theta }{2}\sigma ^{2}}e^{i\frac{\varphi }{2}\sigma ^{3}} 
\rightarrow  S_{+}(n.\sigma )S_{+}^{\dagger}=\sigma ^{3},
\label{S+}
\end{equation}
Then the $A_{4}$ component of the superposition of two L dyons is,
\begin{equation}
  A_{4}^{LL}=\left(-\nu'+\frac{1}{r}\right)\frac{\sigma ^{3}}{2}=\left(-2\pi T+\nu +\frac{1}{r}\right)\frac{\sigma ^{3}}{2}, \;\; r\to \infty .
\end{equation}
To write $A_{4}^{LL}$ in the form of $A_{4}^{MM}$ of Eq. (40), one 
can gauge transform $A_{4}^{LL}$ by the time-dependent matrix $
U=exp(-i\pi Tx^{4}\sigma ^{3})$ \cite{Dia2009},
\begin{equation}
  A_{4}^{LL}=(\nu +\frac{1}{r})\frac{\sigma ^{3}}{2}, \;\; r\rightarrow  \infty .
\end{equation}
And the radial component of the electric and magnetic fields of the 
superposition of two L dyons at infinity are,
\begin{equation}
  E_{r}^{LL}=B_{r}^{LL}=-\frac{1}{r^{2}}\frac{\sigma ^{3}}{2}, \;\; r\rightarrow  \infty
\end{equation}
while the $\theta, \varphi $ components of these fields are 
time-dependent inside the cores of dyons, because the matrix $U$ is 
time-dependent. Therefore, the superposition of two L dyons has Abelian 
electric and magnetic charges (-,-) and when we look at the 
superposition of two L dyons from infinity, we see a point-like charge 
with electric and magnetic charges (-,-), just like the original L dyon. 
It may sound surprising, but we should notice that the deformed dyons 
whose superposition is the solution of Y-M equations are not $U(1)$ objects 
at infinity anymore and therefore, they do not have well-defined $U(1)$ electric 
and magnetic charges at large distances. Hence, we cannot discuss about the charges 
of each deformed dyons. On the other hand, their superposition is the 
solution of Y-M equation, thus $SU(2)$ gauge group breaks to $U(1)$ at 
infinity, and then its electric and magnetic charges are well-defined at 
infinity. We should also notice that the interaction of these two 
deformed dyons is not zero even at infinity, and $H_{\mu \nu }^{a}$ 
that defines this interaction, has its own part of electric and magnetic 
fields and charges.

Since the superposition of two dyons is the solution of Y-M equations, 
we can calculate the topological charge and the action just like Eqs. (\ref{tcharge}) and (\ref{action}) of section 
\ref{sec:dyon},
\begin{equation}
\begin{split}
  Q_{T}&=-\frac{1}{8\pi ^{2}}\int_{0}^{\frac{1}{T}}{dx_{4}\int{d^{3}}rE_{i}^{'a}B_{i}^{'a}} \\{}
 & =-\frac{1}{2\pi T}\int{dr\frac{d}{dr}\frac{(1-A^{'2})E'}{r}}=-\frac{1}{2\pi T}\frac{(1-A^{'2})E'}{r}\vert _{0}^{\infty }=\frac{\nu }{2\pi T}, 
\end{split}
\end{equation}
and 
\begin{equation}
  S=\frac{1}{4g^{2}}\int{d^{4}}x F_{\mu \nu }^{\prime a}F_{\mu \nu}^{\prime a}=\frac{1}{4g^{2}}\int{d^{4}}x F_{\mu \nu }^{\prime a}{\widetilde{F} }_{\mu \nu }^{\prime a}
  =\frac{8\pi ^{2}}{g^{2}}Q_{T}.
\end{equation}
The result is that if the superposition of two dyons is given as a solution of Y-M 
equation, the system behaves just like one original individual dyon at infinity. 
However, this superposition inside the cores of dyons loses the 
spherical symmetry and the gauge fields of Eqs. (\ref{new-A4}) and (\ref{new-Ai}) 
depends on $l$, the separation of the centers of dyons and polar 
angle $\theta $, shown in Fig. (\ref{fig:2dyons}) \cite{KiaDel2018}.

\section{The superposition of M and L dyon}
\label{sec:ML}

Finally, we study the superposition of a pair of M and L dyons. We 
repeat all we did in sections \ref{sec:superposition} and \ref{sec:MM,LL}, but it should be noted that the 
second dyon is L dyon and its profile functions should be different from 
the profile functions of M dyon. Therefore, from Eqs. (\ref{app-new1}) and (\ref{app-new2}), the 
equations deforming the dyon M due to the presence of the dyon L are,
\begin{equation}
  A^{(M)}(r)\frac{E^{(M)}(r)}{r}-\frac{d}{dr}A^{(M)}(r)\approx (1-A^{(M)}(r))\frac{E^{(L)}(r)}{r},
  \label{new1-ML}
\end{equation}
\begin{equation}
  r\frac{d}{dr}\left(\frac{E^{(M)}(r)}{r}\right)-\frac{A^{(M)2}(r)-1}{r}\approx (1-A^{(M)}(r))\frac{1-A^{(L)}(r)}{r},
  \label{new2-ML}
\end{equation}
where $E^{(M)}(r)$ and $A^{(M)}(r)$ are the profile functions of the first dyon M,
\begin{equation}
  E^{(M)}(r)=ar+b,\;\;\;\; A^{(M)}(r)=c,
\end{equation}
while, $E^{(L)}(r)$ and $A^{(L)}(r)$ are the profile functions of the second dyon L,
\begin{equation}
  E^{(L)}(r)=a'r+b',\;\;\;\; A^{(L)}(r)=c'.
\end{equation}
Putting these function to Eqs. (\ref{new1-ML}) and (\ref{new2-ML}),
\begin{equation}
  c\left(a+\frac{b}{r}\right)=(1-c)\left(a'+\frac{b'}{r}\right),
\end{equation}
\begin{equation}
  -\frac{b}{r}-\frac{c^{2}-1}{r}=(1-c)\left(\frac{1-c'}{r}\right),
\end{equation}
Solving the above simple equations for $r^{0}$ and $\frac{1}{r}$ terms, 
the following equations are found,
\begin{equation}
  ca=(1-c)a',\;\;\; cb=(1-c)b',\;\;\; -b-c^{2}=cc'-c-c'.
  \label{abc}
\end{equation}
Changing the role of M and L dyons to study the effect of the presence of dyon M on 
dyon L in Eqs. (\ref{new1-ML}) and (\ref{new2-ML}), three other equations are found,
\begin{equation}
  c'a'=(1-c')a,\;\;\; c'b'=(1-c')b,\;\;\; -b'-c^{\prime 2}=cc'-c-c'.
  \label{abc'}
\end{equation}
Solving these equations, we find
\begin{equation}
  b+b'=1,\;\;\;\; c+c'=1,
  \label{bc}
\end{equation}
while, $a+a'$ remains unspecified. To check the boundary conditions, 
we calculate the electric field as done in section \ref{sec:superposition},
\begin{equation}
  E_{i}^{(1)a}=(\delta _{ai}-n_{a}n_{i})\frac{A^{(M)}(r)}{r}\frac{E^{(M)}(r)}{r}+n_{a}n_{i}\frac{d}{dr}\left(\frac{E^{(M)}(r)}{r}\right),
\end{equation}
\begin{equation}
  E_{i}^{(2)a}=(\delta _{ai}-n_{a}n_{i})\frac{A^{(L)}(r)}{r}\frac{E^{(L)}(r)}{r}+n_{a}n_{i}\frac{d}{dr}\left(\frac{E^{(L)}(r)}{r}\right),
\end{equation}
\begin{equation}
  H_{i4}^{a}=-(\delta _{ai}-n_{a}n_{i})\left[ 
\frac{1-A^{(M)}(r)}{r}\frac{E^{(L)}(r)}{r}+\frac{1-A^{(L)}(r)}{r}\frac{E^{(M)}(r)}{r}\right]
\end{equation}
then
\begin{equation}
\begin{split}
  E_{i}^{'a}&=(\delta _{ai}-n_{a}n_{i})\left[ 
\frac{c}{r}\left(a+\frac{b}{r}\right)+\frac{c'}{r}\left(a'+\frac{b'}{r}\right)-\frac{1-c}{r}\left(a^{'}+\frac{b^{'}}{r}\right)-\frac{1-c'}{r}\left(a+\frac{b}{r}\right)\right] \\{}
&+n_{a}n_{i}\left(-\frac{b}{r^{2}}-\frac{b'}{r^{2}}\right),
\end{split}
\end{equation}
The coefficients of $\frac{1}{r}$ and $\frac{1}{r^{2}}$ terms in the 
first part of the electric field are $
(ca+c'a'-(1-c)a'+(1-c')a ) $and $(cb+c'b'-(1-c)b'+(1-c')b)$, respectively, where both of 
them are zero by Eqs. (\ref{abc}) and (\ref{abc'}). Hence the electric field of the 
superposition of M and L dyons is,
\begin{equation}
  E_{i}^{'a}=-n_{a}n_{i}\left(\frac{b+b'}{r^{2}}\right)=n_{a}n_{i}\left(\frac{-1}{r^{2}}\right),
\end{equation}
where, $b+b'=1$ by Eq. (\ref{bc}). Therefore, the superposition of M and 
L dyons justifies the boundary conditions despite the unspecified 
parameters. We can find the gauge fields of this superposition as we did 
in Eqs. (\ref{A'})-(\ref{new-Ai}) of section \ref{sec:MM,LL},
\begin{equation}
  A_{\mu }^{\prime a}(r)=A_{\mu }^{a}(r)+B_{\mu }^{a}(r'),
  \tag{\ref{A'}}
\end{equation}
\begin{equation}
\begin{split}
  A_{4}^{\prime a}(r)&=A_{4}^{a}(r)+B_{4}^{a}(r')\approx n_{a}\left[
(a+a')+\frac{b+b'}{r}\right] =n_{a}\left(a+a'+\frac{1}{r}\right) \\{}
&\rightarrow  E'(r)=(a+a')r+1
\end{split}
\end{equation}
\begin{equation}
  A_{i}^{\prime a}(r)=A_{i}^{a}(r)+B_{i}^{a}(r')\approx \epsilon _{aij}n_{j}\frac{2-(c+c')}{r}
  =\epsilon _{aij}n_{j}\frac{1}{r} \rightarrow A'(r)=0.
\end{equation}
The coefficient $(a+a')$ which is the holonomy of the new solution 
of Y-M equations is the free parameter of our problem. However, since we 
prefer that the two dyons have the same effect on each other and in addition we would like to use the 
holonomies of dyons as the characteristic parameters, we choose $
a=a'=(\nu +\nu')/2$. Thus, $a+a'=\nu +2\pi T-\nu =2\pi T$
. Applying the unitary matrix $S_{+}(\theta, \varphi )$ of Eq. (\ref{S+}) 
and then time-dependent matrix $U=exp(-i\pi Tx^{4}\sigma ^{3})$, one 
can find the gauge rotated $A_{4}^{ML}$ and the electric field at 
infinity,
\begin{equation}
  A_{4}^{ML}=\frac{1}{r}\frac{\sigma ^{3}}{2},\;\; r\rightarrow  \infty .
\end{equation}
\begin{equation}
  E_{r}^{ML}=B_{r}^{ML}=-\frac{1}{r^{2}}\frac{\sigma ^{3}}{2},\;\; r\rightarrow \infty
\end{equation}
Therefore, the superposition of M and L dyons as the solution of Y-M 
equation is a $U(1)$ object with the electric and the magnetic charges 
(-,-) and trivial holonomy at infinity. Hence, this superposition has no 
contribution in the confinement phase.

\section{Conclusion}
\label{sec:concl}

The profile functions of two overlapped dyons is studied at infinity. Since the superposition of two or more dyons is not the solution of Y-M equation, dyons are often considered in stringy gauge so that the sum ansatz of their gauge fields are as close as possible to the solution of Y-M equations, at least at infinity. 
Unlike this common procedure, we assume that the superposition of two dyons is the solution of Y-M equations and then we find new version of the Y-M equations for individual dyons which apparently do not satisfy the original equations, anymore. In other word, we assume that when two dyons get close to each other, they deform themselves in such a way that their superposition becomes the solution of Y-M equations. Therefore, we do not need to limit ourselves to the stringy gauge and we do our calculations in the hedgehog gauge, in which dyons are originally defined. 
In our previous paper \cite{KiaDel2018}, we used this method to study the interaction of two dyons in the region of their cores called near field limit. We showed 
that in this case, the profile functions of dyons depends on the polar angle and the spherical symmetry is lost. 
 
In this paper, we study the profile functions of two dyons in the far field limit. We solve the Y-M equations associated to the two dyonic system at infinity with the boundary conditions which keep the action finite. It is shown that the gauge fields of this superposition at infinity, are just like an original individual dyon. Then, we calculate the electric and magnetic fields and their charges and show that the superposition of two M(L) dyons has the Abelian  +1(-1) electric and +1(-1) magnetic charges, just like one M(L) dyon.
We also consider the superposition of two different type of dyons, M and L dyons, as the solution of Y-M equation. We find out that, with our special choice of holonomies, this new structure has the electric and magnetic charges (-,-) and its holonomy is trivial and this structure has no role in studying the confinement phase. We recall that, this structure is completely different compared with calorons with M and L dyons and an extra twist on one of the dyons.

\section{Acknowledgement}
 
We are grateful to the research council of the University of Tehran and Iran National Science Foundation: INSF for supporting this study.

\end{document}